# Optimal Placement of Detectors to Minimize Casualties on a Manmade Attack

**Mohammad Marufuzzaman**[1]**, Amin Aghalari**[1]**, Randy Buchanan**[2]**,
Christina H. Rinaudo**[2]**, Kayla M. Houte**[2]**, Julekha H. Ranta**[1]

[1]Department of Industrial and Systems Engineering, Mississippi State University, MS, 39762
[2]U.S. Army Engineer Research and Development Center, Vicksburg, MS, 39180

**Abstract**

This study proposes a mathematical model to optimally locate a set of detectors in such a way that the expected number of casualties in a given threat area can be minimized. Detectors may not be perfectly reliable, which is often a function of how long an attacker would stay within the detectors effective detection radius. To accurately detect any threat event and to avoid any false alarm, we assume that a set of backup/secondary detectors are available to support the primary detectors. The problem is formulated as a nonlinear binary integer programming model and then solved as a linearized branch-and-bound algorithm. A number of sensitivity analyses are performed to illustrate the robustness of the model and to draw key managerial insights. Experimental results reveal that a two-layer detection will significantly minimize the expected number of casualties in a threat area over a one-layer detection problem.

## 1.Introduction

In the past few decades there has been a significant and clear growing trend towards the number of manmade attacks (e.g., active shooting, bomb threat) in the United States and all over the world. For instance, The Federal Bureau of Investigation (FBI) reports that the average number of active shooting incidents between 2000-2007 was 7.4 incidents/year. This number increased to 17.6 incidents/year between 2008-2015, and finally accelerated to 25.7 incidents/year between 2016-2017 [1]. Most importantly, majority of such attacks are occurring in the gun free zones (e.g., education institutes, businesses and shopping malls, hospitals), include all areas where the general public is forbidden to carry firearms. For instance, 90% of the past active shooting incidents occur in different gun free zones, including Pre-K to 12 schools (14.8%), universities (6%), businesses and shopping malls (42%) to even in health care facilities (4%). Such violence in the gun free zones poses serious security concerns among public safety, primarily due to the horrifying outcomes and potentially large number of casualties that typically stem from such an attack. All the above underlying statistics indicate that there is an urgent need to methodologically securing a workplace not only to minimize the expected number of casualties from a given attack but also to restrain civilians experiencing such rare but dreadful events.

Despite knowing the horrifying outcomes that typically stem from such a pre-planned attack, our knowledge on how best a workplace (or even a threat area) can be secured is still very limited. A majority of the design decisions (e.g., optimal placement of detec-tors for detecting any threat element (guns, bombs)) to secure a workplace are primarily made based on intuition rather adopting a methodological framework. Accordingly, very limited articles are available in the literature that research on this pressing problem. The first novel initiative along this line of research probably was conducted by Kaplan and Kress [2] who analyze the operational effectiveness and consequences among individu-als under a suicide bomber detection scheme. Given the detectors may not be perfectly reliable, Nie et al., [3] extended this work by considering the reliability of a single type of detectors. Later, Yan and Nie [4] extended this framework to consider multiple types of detectors with an application of vessel attack in a maritime port.

Besides modeling approaches, few researchers adopted discrete event simulation approach (e.g., citeLee2019) to assess the workplace safety under an active shooting situation. Note that in these studies the researchers indirectly assessed the workplace safety by examined the civilians responses under an active shooting situation. Different than the studies dis-cussed above, specially in [3] and [4], our study assumes that a set of secondary/backup detectors are available in a workplace or any threat area in general to support the primary detectors. The purpose of these secondary/backup detectors are twofold: increasing the accuracy in detecting any (i) threat event to minimize the expected number of casualties from an attack and (ii) false alarm to unnecessarily confuse/shock both the civilians and the law enforcement agencies. In overall, we propose a nonlinear binary integer program-ming model to optimally locate a set of detectors (both primary and secondary) in such a way that the expected number of casualties from an unexpected attack can be minimized.

## 2. Problem Description and Mathematical Model Formulation

The purpose of this study is to optimally deploy a set of primary and secondary detec-tors on a threat area in such a way that the expected number of casualties from a sudden attack can be minimized. We assume that both the primary and secondary detectors are stationary and perfectly concealed. We further assume that the interdiction team will start preparing for the mission after receiving an alarm from the primary detectors while the purpose of the secondary detectors is to check any *false detection* before the mission initiates. The goal would be to locate the detectors (both primary and secondary) in such a way that both the attacker(s) and threat element(s) (e.g., bomb, gun) are appropriately detected and the interdiction team will receive a *minimum response time* to baffle the attacker's intention. To formulate this problem, we first assume that the threat area has been divided into $m \times n$ grids and is denoted by set $G = \{j: j = 1, 2, \dots, mn\}$. In this grid the set associated with entrance of attacker, blocked grids, unblocked grids, and the potential target of attacker is presented with $E, B, U$, and $A$ respectively. We further let $\gamma_{ej}$ to be the probability associated with using an entrance $e \in E$ to attack threat grid $j \in A$ in a given time period. Hence, $\gamma_{ej} \geq 0; \forall e = 1, 2, \dots, E$ and $\sum_{e=1}^{E} \sum_{j \in A} \gamma_{ej} = 1$. It is reasonable to assume that an attacker will enter into the threat area from an entrance $e \in E$ and then use a shortest path, denoted b $d_{ej}$, to reach a threat grid $j \in A$, if not obstructed by any of the blocked grids $j \in B$. To find $d_{ej}$, we apply Dijkstra's algorithm [5], and we assume that the attacker will walk, with a speed of $k$ m/sec, from one grid center to the next till she/he reaches to a threat grid $j \in A$.

In this study we consider two types of detectors: *primary* and *secondary* detectors. Let $\alpha^p, \beta^p$, and $\psi^p$ be the effective detection radius, instantaneous detection rate, and unit purchasing cost for primary detectors. To increase the detection probability and to minimize the false detection rate, we assume that an additional layer of protection, in the form of secondary/backup detectors (e.g., radar and video together, thermal imagery, terahertz radar), can be placed in selected critical grids. Let $\alpha^s$, $\beta^s$, and $s$ be the effective detection radius, instantaneous detection rate, and unit purchasing cost for secondary detectors. We let $M^p$ and $M^s$ to denote the availability of budget to procure primary and secondary detectors, respectively. Preventing any threat event usually consists of three main steps: (i) detection, (ii) response, and (iii) neutralization. The placement of the detectors should be made in such a way that the interdiction team receives at least $\chi^*$ seconds to neutralize the threat. Hence, given the walking speed of the attacker be $k$ m/sec, this converts the threat detection to be made at least $k\chi^*$ meter away (e.g., $10\ m$ as defined by Kaplan and Kress [2]). Let $N_e^p(j)$ be the set of all non-blocked grids on which a primary detector can be located and timely detect the attacker while traveling on path $d_{ej}^2$. We assume that as soon as the primary detector is able to detect any threat, the interdiction team will start preparing for the neutralization process i.e., the response step starts. However, the actual neutralization process will only start after receiving a confirmation alarm (i.e., no false detection) from the secondary detectors. Let $N_e^s$ be the set of all non-blocked grids on which a secondary detector can be located and confirm any threat event, along with the attacker, while the attacker may travel on path $d_{ej}$.

We assume that the detectors (both primary and secondary detectors) are not perfectly reliable. There is always a probability that the detectors are unable to detect any threat which is essentially a function of the duration an attacker stay within the effective detec-tion radius of the detectors. For every $i \in N_e^p(j)$, let $\rho_{iej}$ be the probability associated with detecting any threat event by the primary detector which has been carried out by the attacker while traveling on path $d_{ej}$.

Likewise, for every $i' \in N_e^s(j); \forall (i, i') \in U, i \neq i'$, let $\rho_{i'ej}$ be the probability associated with detecting any threat event by the secondary detector which has been carried out by the attacker while traveling on path $d_{ej}$. Based on Przemieniecki [6], we calculate $\rho_{iej}$ and $\rho_{i'ej}$ as follows: $\rho_{iej} = 1 - e^{-\beta^p l_{iej}^p}$ and $\rho_{i'ej} = 1 - e^{-\beta^s l_{i'ej}^s}$ where $\beta^p$ and $\beta^s$, respectively, denote the instantaneous detection rate for the primary and secondary detectors. Further, parameters $l_{iej}^p$ and $l_{i'ej'}^s$, respectively, denote the portion of the travel length on path $d_{ej}$ while the attacker is within the effective detection radius of primary and secondary detectors. A procedure to geometrically calculate $l_{iej}^p$ can be found in [3] and [4]. We define sets $P = \cup_{e,j} N_e^p(j)$ and $S = \cup_{e,j} N_e^s(j)$ which will realistically restrict the detectors (primary and secondary) placement decisions in our tested threat region. Let $X \coloneqq \{X_j\}_{\forall j \in P}$ and $Y \coloneqq \{Y_j\}_{\forall j \in S}$ to denote binary decision variables if a detector (primary or secondary) is placed on grid $j \in P \cup S$.

All possible events that may essentially incur to any threat event are illustrated in Figure 1. We assume that the two detectors are independently correlated with each other. Figure 1 clearly depicts that there are four possible scenarios where an attack via path $d_{ej}$ may turns out to be successful (nodes 3, 5, 7, and 9 in Figure 1). Eventually, we assume that $\theta_1$ and $\theta_2$ are the probability of success of any neutralization associated with node 7 and 9, respectively.

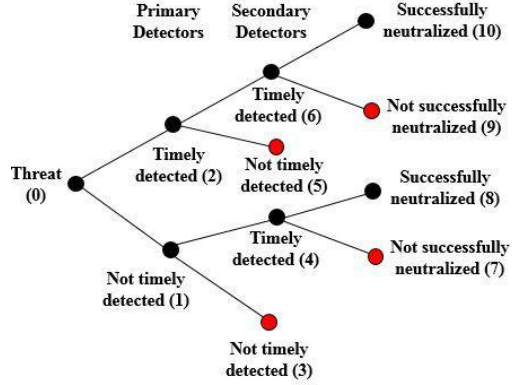

Figure 1: All possible events related to a threat

Let $\zeta_j$ to denote the expected number of casualties at attack grid $j \in A$. Given that an attacker utilizes path $d_{ej}$ with probability $\gamma_{ej}$, and also rearranging the total expected number of casualties for four possible scenarios, we introduce the following nonlinear binary integer program for our optimal detector placement problem.

$$[DP]\ Minimize \sum_{e=1}^{E}\sum_{j\in A}[(\theta_1-\theta_2)\prod_{i\in N_e^p(j)}(1-\rho_{iej})^{X_i}\zeta_{ej}+\theta_1\prod_{i'\in N_e^s(j)}(1-\rho_{i'ej})^{Y_{i'}}\zeta_j\gamma_{ej}$$
$$-(\theta_1-\theta_2)\prod_{i\in N_e^p(j)}(1-\rho_{iej})^{X_j}\prod_{i'\in N_e^s(j)}(1-\rho_{i'ej})^{Y_{i'}}\zeta_j\gamma_{ej}]$$

Subject to

$$\sum_{j\in P}\psi^p X_j \leq M^p \qquad (1)$$

$$\sum_{j\in S}\psi^s Y_j \leq M^s \qquad (2)$$

$$X_j + Y_j \leq 1 \qquad \forall j \in P \cap S \qquad (3)$$

$$Y_j \leq \sum_{j'\in P} X_{j'} \qquad \forall j \in S \qquad (4)$$

$$X_j \in \{0,1\} \qquad \forall j \in P \qquad (5)$$

$$Y_j \in \{0,1\} \qquad \forall j \in S \qquad (6)$$

The objective of model **[DP]** is to minimize the expected number of casualties due to a sudden manmade attack. Constraints (1) and (2) ensure that the cost of placing detectors (primary and secondary) is restricted to their budget availability. Constraints (3) ensure that at most one detector, either primary or secondary, can be placed in a given grid 2 P $^T$ S. Constraints (4) ensure that a secondary detector may support multiple primary detectors. Finally, constraints (5) and (6) set binary restrictions for the detectors placement decisions.

## 3. Branch and Bound Algorithm

Model **[DP]** is nonlinear due to containing several nonlinear terms in the objective func-tion. To linearize model **[DP]** and to provide quality solutions in a reasonable time, we use the linearized branch and bound algorithm proposed by Yan and Nie [4] which is extended over the study from Sherali et al. [7] to solve a nonconvex continuous optimiza-tion model. Further, it is worth mentioning that having applied linearization technique proposed by Yan and Nie [4], there will be a bilinear term in the objective function to alle-viate which we employ a tighter piecewise McCormick relaxation technique as proposed by Castillo et al. [8]. Detailed information regarding this method could be found in [9].

## 4. Case Study

In this section, results of two experiments conducted on a base case is presented. The first set of experiments study the impact of detection radius for the primary ($a^p$) and sec-ondary ($a^s$) detectors on the expected number of casualties. As evidenced from Figure 2(a), as $a^p$ and $a^s$ increases, the expected number of casualties decreases. Further, the re-sults clearly demonstrate an additional benefit of utilizing a two-layer detection over a one-layer detection mechanism. The next set of experiments study the impact of instantaneous detection rate for primary ($b^p$) and secondary ($b^s$) detectors on the expected number of casualties. Here, it can be seen that as the values of $b^p$ and $b^s$ increases, the number of expected casualties drops. It can be observed from Figure 2(b) that on average an addi-tional 15.6% casualties can be saved if a two-layer detection mechanism is employed over a one-layer detection mechanism.

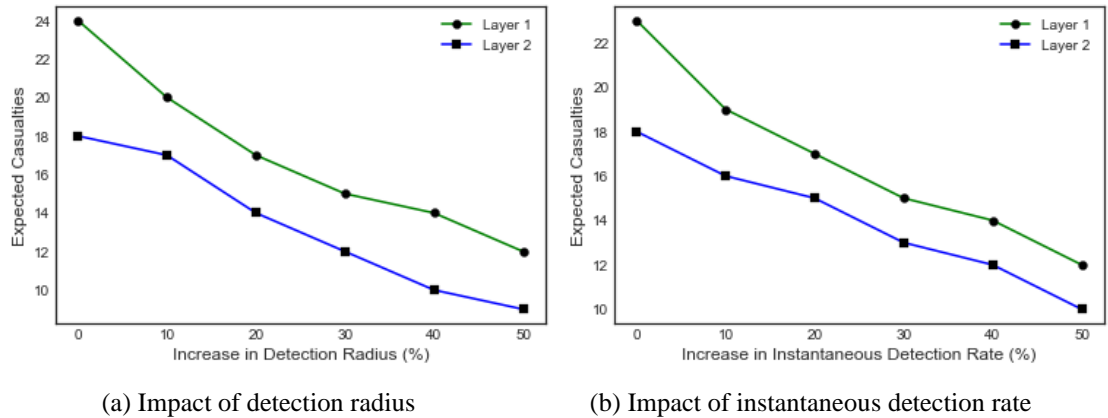

(a) Impact of detection radius          (b) Impact of instantaneous detection rate

Figure 2: Detector placement for two-and-one layer detection problems

## 5. Conclusion

This study proposes an innovative nonlinear binary integer programming model to optimally deploy a set of primary and secondary detectors on a threat area in such a way that the expected number of casualties can be minimized. We then employ a linearized branch-and-bound algorithm to solve our proposed mathematical model. To the end, a number of sensitivity analyses are performed to illustrate the robustness of the model and to draw key managerial insights. In our future research, we will examine how the model would behave under a dynamic threat attack [10]. Further, In this article, we test our model under deterministic input parameters. It would be interesting to examine how this model behaves under stochastic parameters and larger environments (e.g., shopping malls, hospitals, etc.) [11, 12].

## Acknowledgements

The authors gratefully acknowledge the funding support from U.S. Army Engineer Re-search and Development Center (ERDC).

## References

[1] Federal Bureau of Investigation "FBI".Quick Look: 250 active shooter incidents in the United States from 2000 to 2017. Available from: https://www.fbi.gov/about/partnerships/office-of-partner-engagement/ active-shooter-incidents-graphics, 2018.

[2] E.H. Kaplan, M. Kress. Operational effectiveness of suicide-bomber-detector schemes: A best-case analysis. Proceedings of National Academy of Sciences of the USA, 102(29):10399–10404, 2005.

[3] X. Nie, R. Batta, C. Drury, L. Lin. Optimal placement of suicide bomber detectors. Military Operations Research, 12(2):65–78, 2007.

[4] X. Yan, X. Nie. Optimal placement of multiple types of detectors under a small vessel attack threat to port security. Transportation Research Part E, 93:71–94, 2016.

[5] E. W. Dijkstra. A note on two problems in connexion with graphs. Numerische Nathe-matik, 1(1):269–271, 1959.

[6] J.S. Przemieniecki. Mathematical methods in defense analyses. American Institute of Aeronautics and Astronautics, Reston, VA, 2000.

[7] H. D. Sherali, J. Desai, T. S. Glickman. Allocating emergency response resources to minimize risk with equity considerations. American Journal of Mathematical and Man-agement Sciences, 24(3–4):367–410, 2004.

[8] P. Castillo, P. M. Castro, V. Mahalec. Global optimization algorithm for large-scale refinery planning models with bilinear terms. Industrial & Engineering Chemistry Re-search, 56:530–548, 2007.

[9] M. Marufuzamman, A. Aghalari, R. K. Buchanan, C. H. Rinaudo, K. M. Houte, J. H. Ranta. Optimal Placement of Detectors to Minimize Casualties in an Intentional Attack. *IEEE Transactions on Engineering Management,* 2020.

[10] M. Marufuzzaman, A. Aghalari, J. H. Ranta, R. Jaradat, "Optimizing Civilian Response Strategy Under an Active Shooting Incident," in *IEEE Systems Journal*, doi: 10.1109/JSYST.2020.3041376.

[11] A. Aghalari, F. Nur, M. Marufuzzaman. A Bender's based nested decomposition algorithm to solve a stochastic inland waterway port management problem considering perishable product. International Journal of Production Economics, 229, 107863, 2020.

[12] A. Aghalari, F. Nur, M. Marufuzzaman. Solving a stochastic inland waterway port management problem using a parallelized hybrid decomposition algorithm. Omega, 102316, 2020.